\documentclass[conference]{IEEEtran}
\IEEEoverridecommandlockouts 

\ifCLASSINFOpdf
\else
\fi

\usepackage{graphicx}
\usepackage{subcaption}
\usepackage{amsmath}
\usepackage{amssymb}
\usepackage{xcolor}
\usepackage{algorithm}
\usepackage{algorithmic}
\usepackage{cuted}
\usepackage{afterpage}
\usepackage{tabularx}
\usepackage{array}
\usepackage{booktabs}
\usepackage{ragged2e}
\usepackage{amsthm}
\usepackage{url}

\allowdisplaybreaks[2]

\begin{document}

\title{ImCDN: An Immersive Content Delivery Network for Cooperative Tile-level 360$^\circ$ Video Streaming%
\thanks{This work was supported in part by the U.S. National Science Foundation under Grant CNS-2225427.}
}

\author{
\IEEEauthorblockN{
Ning Wang\IEEEauthorrefmark{1},
Xiaofan Liu\IEEEauthorrefmark{1},
Beatriz Lorenzo\IEEEauthorrefmark{1}, and
Sumudu Samarakoon\IEEEauthorrefmark{2}
}
\IEEEauthorblockA{\IEEEauthorrefmark{1}
Department of Electrical and Computer Engineering,
University of Massachusetts Amherst,
Amherst, MA, USA\\
}
\IEEEauthorblockA{\IEEEauthorrefmark{2}
Centre for Wireless Communications (CWC),
University of Oulu,
Oulu, Finland\\
Emails: \{ninwang, xiaofanl, blorenzo\}@umass.edu, sumudu.samarakoon@oulu.fi
}
}

\maketitle

\begin{abstract}
Delivering immersive 360$^\circ$ video over Content Distribution Networks (CDNs) demands ultra-high bitrates, yet each user's viewport covers only a small fraction of the full panorama. Therefore, conventional segment-level architectures waste scarce edge resources on unseen regions. This paper presents \textbf{ImCDN}, an immersive end-to-end tile-granularity CDN framework that integrates field-of-view (FoV)-driven cloud pre-distribution with hybrid segment–tile caching at edge servers, edge compute for on-demand tile transcoding, and an edge-to-edge relay mechanism for cooperative content sharing. We formulate the CDN optimization problem, comprising cache admission, compute scheduling, and serving-node selection, as a long-term utility maximization that balances cache hit-rate, edge sharing, end-to-end latency, backhaul traffic, energy consumption, viewport video quality, and cache replacement cost. We solve it via multi-agent proximal policy optimization (MAPPO) under centralized training and decentralized execution across edge servers. Simulations on a realistic topology with real 360$^\circ$ head-tracking traces show that ImCDN improves edge hit rate and reduces user waiting time, reduces origin traffic by approximately 18\% compared to the non-cooperative edge variant, and outperforms segment-only CDN baselines at comparable immersive quality (e.g., 4K/6K) and standard caching heuristics.
\end{abstract}








\IEEEpeerreviewmaketitle

\section{Introduction}
\label{sec:intro}

The rapid adoption of head-mounted displays has driven explosive growth in 360$^\circ$ video traffic, yet delivering such content remains inefficient~\cite{grandview_xr2024}.
A full spherical panorama in equirectangular projection requires ultra-high bitrates (e.g., 6K--8K), whereas each user's FoV spans only about 100$^\circ\times100^\circ$, i.e., less than one-sixth of the sphere~\cite{corbillon2017viewport,xu2018gaze}.
Conventional CDNs cache and deliver each temporal video segment of the full panorama at a single bitrate, wasting scarce bandwidth and edge storage on regions of the panorama that fall outside the user’s viewport~\cite{pang2022towards}.
This mismatch between full-panorama delivery and viewport-centric perception motivates \emph{tile-level}, FoV-aware caching and delivery across the CDN.

Prior research has progressed along several complementary directions, spanning viewport-driven bit rate adaptation at clients~\cite{xie2017360probdash}, edge caching schemes~\cite{LI2025111461}, and learning-based control for CDN optimization~\cite{Malektaji2021DRL_EdgeCDN}.  
Tile-based streaming divides each 360$^\circ$ frame into a spatial grid, allowing clients to request only tiles within the predicted viewport~\cite{xie2017360probdash}. 
However, most studies focus on client-side adaptive bitrate (ABR) design~\cite{zhang2019drl360}, with limited attention to network-side caching and delivery. 
Proactive edge caching has been widely studied for conventional video, ranging from popularity-based heuristics such as Least Recently Used (LRU)
and Least Frequently Used (LFU) to more recent cache-reuse policies for similar semantic content using lifelong reinforcement learning~\cite{11396413}.  
Recent work has extended edge caching to 360$^\circ$ video~\cite{maniotis2022tile,liu2023viewport,LI2025111461}, but most schemes still cache at segment granularity ~\cite{maniotis2022tile,liu2023viewport} or focus on single-edge delivery without exploiting multi-edge content distribution for multi-granularity caching~\cite{LI2025111461}. 

Recent edge computing studies have explored tile-based 360$^\circ$ video delivery. Tan et al.~\cite{10271717} jointly optimize multi-user tile delivery and bitrate allocation under bandwidth and deadline constraints, but focus on a single edge server and assume tiles are served either locally or from the origin. Zeng et al.~\cite{11393688} jointly optimize edge caching and tile bitrate selection for fairness-aware streaming, but do not consider multi-granularity caching, inter-edge cooperation, or CDN-level routing.
\textcolor{black}{Reinforcement learning has been applied to CDN caching optimization,
from adaptive content placement in hierarchical CDNs~\cite{Sadeghi2019DRL_CDN}
to cooperative content migration among edge nodes~\cite{Malektaji2021DRL_EdgeCDN}.
However, existing RL-based CDN solutions treat content as monolithic segments
and do not exploit the tile-level spatial structure of immersive video
or coordinate multi-granularity caching with edge compute and relay mechanisms.}

This paper addresses these gaps by presenting ImCDN (\emph{Immersive CDN}), an end-to-end framework for tile-level 360$^\circ$ video delivery coordinated by distributed edge learning.
The main contributions are as follows:
\begin{enumerate}
\item We propose a tile-aware hierarchical CDN architecture comprising (i)~a cloud layer that performs FoV-driven Top/Near tile classification and proactive edge pre-distribution, (ii)~an edge layer equipped with hybrid caching (segment and tile cache) and edge compute for on-demand tile transcoding, and (iii)~an edge-to-edge cooperative content sharing mechanism among neighboring edges.
\item We formulate the CDN optimization problem as a utility maximization that jointly considers cache hit, edge sharing, latency, backhaul traffic, transmission and compute energy, viewport video quality, and cache replacement cost.
To enable real-time decentralized control, we reformulate the problem as a Markov Decision Process with compact observation and continuous action spaces, and solve it via multi-agent proximal policy optimization (MAPPO) under a centralized-training, decentralized-execution paradigm.
\item We conduct comprehensive simulations on a realistic CDN topology with traced 360$^\circ$ head orientations, and demonstrate that ImCDN achieves higher hit rates, lower user waiting times, and substantially reduced origin traffic compared to both conventional segment-only CDN designs and standard caching heuristics, while edge cooperation nearly doubles the reward relative to the non-cooperative variant.
\end{enumerate}


\section{\textcolor{black}{System Model}}
\label{sec:system_model}


\subsection{CDN Architecture and Content Model}
We consider a hierarchical CDN composed of a cloud (origin) layer, an edge layer, and end users.
Let $\mathcal{O}$ be the set of cloud/origin servers,
$\mathcal{E}$ the set of edge servers equipped with caching and computing resources,
and $\mathcal{U}$ the set of users.
The physical infrastructure may contain intermediate \emph{relay} nodes
(without caching/compute) that only forward traffic; we treat them as part of the underlay network.
We define the set of potential serving nodes as $\mathcal{N}=\mathcal{O}\cup\mathcal{E}$.
The logical connectivity is modeled by a directed overlay graph
$\mathcal{G}^{\mathrm{over}}=(\Omega,\mathcal{A})$ with
$\Omega=\mathcal{O}\cup\mathcal{E}\cup\mathcal{U}$,
where $\mathcal{A}$ is the set of directed arcs representing
all feasible content transfers among overlay nodes.

Let $\mathcal{V}$ denote the set of 360$^\circ$ videos.
Each video $v\in\mathcal{V}$ is partitioned into $N_v$ temporal segments of duration $\Delta t_{\mathrm{seg}}$, indexed by $\mathcal{S}_v=\{1,\dots,N_v\}$.
We model the video using an equirectangular tiled representation with an $n_{\mathrm{cols}}\times n_{\mathrm{rows}}$
grid and tile index set $\mathcal{I}=\{1,\dots,n_{\mathrm{cols}}n_{\mathrm{rows}}\}$.
For each segment, content is organized into (i) a \emph{base segment} (required for playback), (ii) a set of \emph{Top} enhancement tiles (high viewport probability), and (iii) a set of \emph{Near} enhancement tiles (spatial neighbors of Top tiles); let $\mathcal{L}=\{\textsf{Top},\textsf{Near}\}$ denote the enhancement classes. A base segment is indexed by $m=(v,s)$ with size $b_m$, and an enhancement tile item is indexed as $j=(v,s,i,\ell)$ with $i\in\mathcal{I}$, $\ell\in\mathcal{L}$, and size $b_j$. Defining $\mathcal{M}=\{(v,s)\mid v\in\mathcal{V},\, s\in\mathcal{S}_v\}$ and $\mathcal{J}=\{(v,s,i,\ell)\mid v\in\mathcal{V},\, s\in\mathcal{S}_v,\, i\in\mathcal{I},\, \ell\in\mathcal{L}\}$,
we unify all contents as $\mathcal{C}=\mathcal{M}\cup\mathcal{J}$, where each $c\in\mathcal{C}$ carries size $b_c$.


\subsection{Cloud Planner: FoV-Aware Cloud Pre-distribution}
The cloud hosts the origin server storing all content items and serves as a centralized controller that determines which items are proactively pushed to edge nodes prior to playback.

\paragraph{FOV-driven tiles}
For each user, the FOV is defined by the instantaneous head orientation and a fixed angular viewport. Visible tiles are obtained by projecting the FOV onto the equirectangular grid, and per-tile view counts are accumulated within each segment window. The Top set contains the $M$ most-viewed tiles, and the Near set is given by their 8-connected neighbors excluding the Top tiles.

\paragraph{\textcolor{black}{Edge-level aggregation and pre-distribution policy}}
\textcolor{black}{
For each edge server $e\in\mathcal{E}$, the cloud collects the per-tile FoV counts from every user currently attached to $e$ and sums them into an edge-level aggregate count vector $\mathbf{c}_e\in\mathbb{R}^{|\mathcal{I}|}$, where each entry $c_{e,i}$ records how often tile $i$ falls within the viewports of $e$'s users during the most recent segment window.
The cloud then classifies tiles for each edge independently.
The $M$ tiles with the highest aggregate counts are selected as the \emph{Top} set $\mathcal{I}^{\text{Top}}_e$; the \emph{Near} set $\mathcal{I}^{\text{Near}}_e$ is formed by taking the 8-connected spatial neighbors of every Top tile and removing the Top tiles themselves.
To avoid pushing tiles for which there is insufficient demand, the cloud normalizes the aggregate counts into a probability distribution $h_e(i)=c_{e,i}/\sum_{i'}c_{e,i'}$ and computes the cumulative preference scores $\rho^{\text{Top}}_e=\sum_{i\in\mathcal{I}^{\text{Top}}_e}h_e(i)$ and $\rho^{\text{Near}}_e=\sum_{i\in\mathcal{I}^{\text{Near}}_e}h_e(i)$.
Top enhancement tiles are pushed to edge $e$ only if $\rho^{\text{Top}}_e\ge\tau_{\text{top}}$, and Near tiles only if $\rho^{\text{Near}}_e\ge\tau_{\text{near}}$, where $\tau_{\text{top}}$ and $\tau_{\text{near}}$ are predefined thresholds.
}


\begin{figure}[t]
    \centering
    \includegraphics[width=1\linewidth]{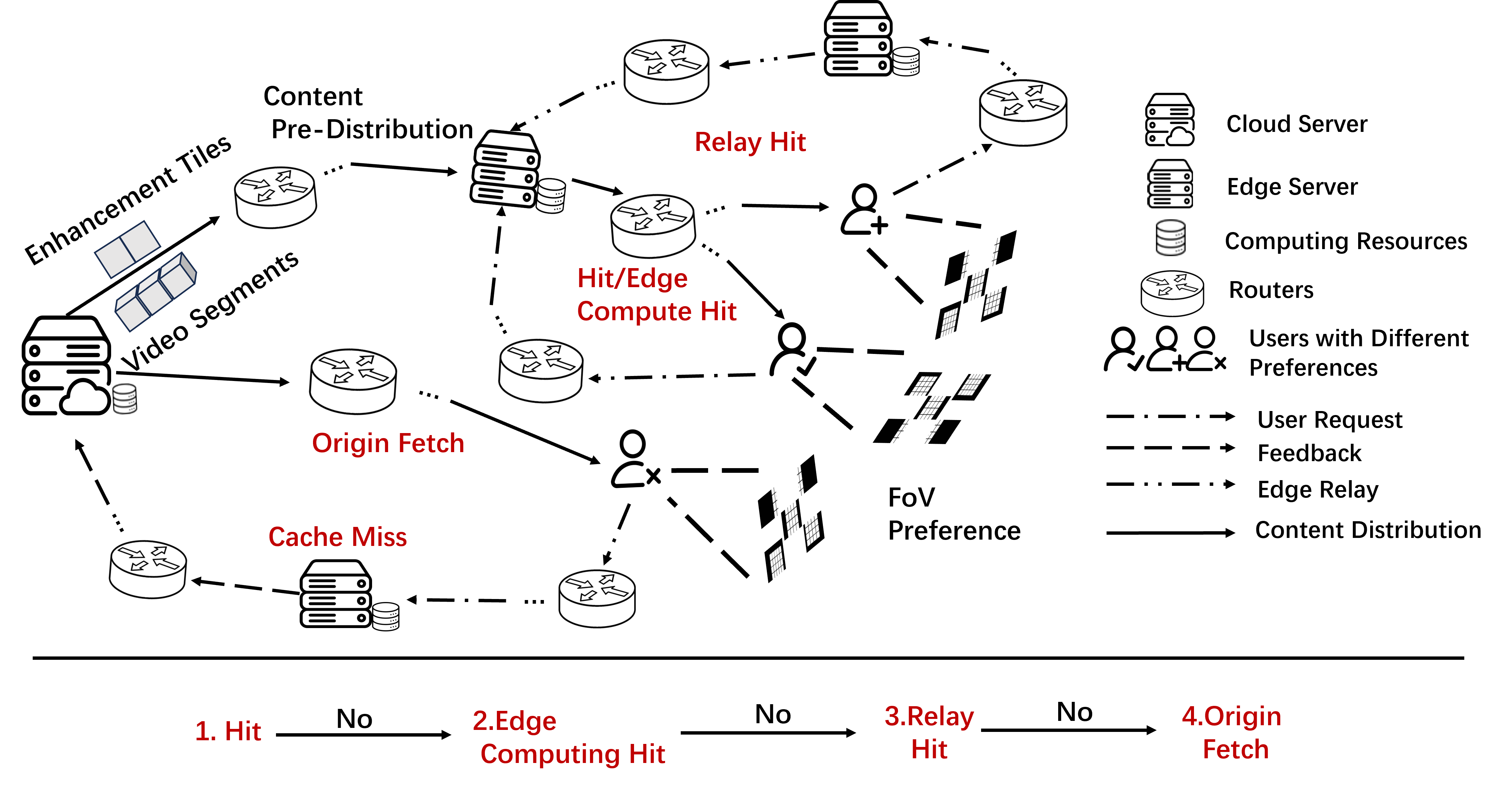}
    \caption{Hierarchical ImCDN Design.}
    \label{fig:CDN network system}
\end{figure}

\subsection{Edge Server Model: Double Caches and Edge Compute}
Each edge server $i\in\mathcal{E}$ is modeled with two logically
separate caches and a bounded compute budget. 

\paragraph{Double-cache architecture and compute budget}
Edge $i$ maintains a segment cache of capacity $B^{\text{seg}}_i$ for base segments $m\in\mathcal{M}$ and a tile cache of capacity $B^{\text{tile}}_i$ for enhancement tiles $j\in\mathcal{J}$. The corresponding binary caching variables are $y_{i,m}\in\{0,1\}$ and $y_{i,j}\in\{0,1\}$, subject to
$\sum_{m\in\mathcal{M}} y_{i,m} b_m \le B^{\text{seg}}_i$ and
$\sum_{j\in\mathcal{J}} y_{i,j} b_j \le B^{\text{tile}}_i$.
In each epoch $\tau$, edge $i$ also has compute budget $C^{\text{comp}}_i(\tau)$, used to generate or adapt enhancement tiles on demand, including generating missing Top/Near tiles from a parent or base representation and downscaling a cached Top tile to serve a Near request. Let $x_{i,j}\in\{0,1\}$ indicate that edge $i$ performs a compute task for tile item $j$, and let $w_j$ denote the corresponding workload; then
$\sum_{j\in\mathcal{J}} w_j x_{i,j} \le C^{\text{comp}}_i(\tau)$.

\paragraph{Serving logic and content sources}
Let $I^s_{u,c,\tau} \in \{0,1\}$ indicate whether the request for content $c$ from user $u$ was served at time $\tau$ by source node $s \in \{hit,e-comp,relay,orig\}$. For each request ($u$, $c$), one serving node is selected at time $\tau$, i.e., $
\sum_{s} I^{s}_{u,c,\tau} \leq 1$.
The edge server resolves each request following the priority order:
1) {\emph{Hit by cache}} (served from cached content, $I^{\text{hit}}_{u,c,\tau}=1$),
2) {\emph{Edge compute hit}} (generated locally within the available compute budget, $I^{\text{e-comp}}_{u,c,\tau}=1$),
3) {\emph{Relay hit}} (retrieved from a neighboring edge server by relaying the request, $I^{\text{relay}}_{u,c,\tau}=1$),
or 4) {\emph{Origin fetch}} (retrieved from the cloud over the backhaul, $I^{\text{orig}}_{u,c,\tau}=1$).
Let $\mathcal{R}(\tau)$ denote the set of requests arriving at time $\tau$.
We define the edge hit rate as
$\mathrm{HR}(\tau)=\frac{\sum_{(u,c)\in\mathcal{R}(\tau)}\!\left(I^{\text{hit}}_{u,c,\tau}+I^{\text{e-comp}}_{u,c,\tau}+I^{\text{relay}}_{u,c,\tau}\right)}{|\mathcal{R}(\tau)|}$,
and the corresponding origin traffic is
$B^{\text{orig}}(\tau)=\sum_{(u,c)\in\mathcal{R}(\tau)} b_c\,I^{\text{orig}}_{u,c,\tau}$.

\subsection{Communication Model}
\label{subsec:comm}
We model the underlying network as a directed graph $\mathcal{G}^{\text{under}}=(\widetilde{\Omega},\widetilde{\mathcal{A}})$, where  $\widetilde{\Omega}$ comprises cloud datacenters, CDN edge servers,
relay nodes, and end users.
Backbone connectivity among cloud/edge/relay nodes is
ensured by construction. Each directed link $e=(u,v)\in\widetilde{\mathcal{A}}$ is characterized by a latency $\delta_e$ and a capacity.
%
Let $d_{u,v}$ be the geographic distance between nodes $u$ and $v$.
Links are partitioned into backbone $\widetilde{\mathcal{A}}^{\text{bb}}$ (cloud/edge/relay) and access $\widetilde{\mathcal{A}}^{\text{ac}}$ (user--relay). Propagation delay is approximated as distance over propagation speed
in optical fiber \cite{kurose2021topdown}.
Latency per link is:
\begin{equation}
\delta_e=
\begin{cases}
\delta_0^{\text{bb}} + \kappa\, d_{u,v}, & e\in\widetilde{\mathcal{A}}^{\text{bb}},\\
\max\!\left\{\delta_{\min}^{\text{ac}},\ \delta_0^{\text{ac}} + \kappa\, d_{u,v} + \xi \right\}, & e\in\widetilde{\mathcal{A}}^{\text{ac}},
\end{cases}
\end{equation}
where $\delta_0^{\text{bb}}$ captures baseline router processing on backbone
links, $\delta_0^{\text{ac}}$ captures last-mile baseline delay, $\delta_{\min}^{\text{ac}}$
prevents unrealistically small access latencies, and $\xi\sim\mathcal{N}(0,\sigma_{\text{ac}}^2)$
models access-layer jitter.  
Each link $e$ has a nominal capacity $C_e$ available to flows (after lower-layer protocol effects and
hardware rate constraints). Backbone links are provisioned at multi-Gbps rates:
$C_e = \alpha_{\mathrm{bw}} \cdot \lambda \cdot X_e, e\in\widetilde{\mathcal{A}}^{\mathrm{bb}}$,
where $X_e\sim\mathcal{U}(X_{\min},X_{\max})$ is the raw link rate in Gbps,
$\lambda$ is the Gbps-to-Mbps conversion factor,
and $\alpha_{\mathrm{bw}}\in(0,1]$ is a global bandwidth scaling factor used for stress testing.
Access links follow tiered broadband rates:
\begin{equation}
\label{eq:ac_capacity}
C_e =
\alpha_{\mathrm{bw}}\;
\mathrm{clip}\!\left(\mathrm{tier}\cdot\zeta,\;
C_{\min}^{\mathrm{ac}},\;C_{\max}^{\mathrm{ac}}\right),
\qquad e\in\widetilde{\mathcal{A}}^{\mathrm{ac}},
\end{equation}
where $\mathrm{clip}(x,a,b)=\min\{b,\max\{a,x\}\}$,
$\mathrm{tier}\in\mathcal{T}_{\mathrm{ac}}$ (Mbps) is sampled from a categorical distribution
over broadband service tiers,
$\zeta\sim\mathcal{U}(\zeta_{\min},\zeta_{\max})$ models access sharing and contention,
and $C_{\min}^{\mathrm{ac}}$, $C_{\max}^{\mathrm{ac}}$ are the minimum and maximum
effective access rates.

We consider a slotted system with slot duration $\Delta t_{\text{seg}}$ (s). In slot $\tau$, link $e$ is modeled in continuous time with transmission budget $\mathrm{cap}_e(\tau)$ under nominal capacity $C_e$, and admitted requests are served in FIFO order. For a request of content $c$ with size $b_c$ bits arriving at time $t_0$, service on link $e$ starts at
$t^{\text{start}}_{e,c}=\max\{t_0,\ t^{\text{finish}}_{e,\mathrm{prev}}\}$,
and finishes at
$t^{\text{finish}}_{e,c}=t^{\text{start}}_{e,c}+\frac{b_c}{\mathrm{cap}_e(\tau)}$,
where $t^{\text{finish}}_{e,\mathrm{prev}}$ is the completion time of the previous request on link $e$. This induces a queuing delay $T^{\text{queue}}_c$. For a delivered content item $c$, the end-to-end completion time is modeled as
\begin{equation}
T^{\text{comp}}_c
=
T^{\text{prop}}_c
+
T^{\text{tx}}_c
+
T^{\text{queue}}_c
+
T^{\text{edge}}_c,
\label{eq:transfer_completion_time}
\end{equation}
where $T^{\text{prop}}_c$ is the sum of per-link propagation latencies along the chosen path,
$T^{\text{tx}}_c$ is the transmission time determined by the data size and capacity,
$T^{\text{queue}}_c$ captures queuing delay under shared link capacities,
and $T^{\text{edge}}_c$ is the edge computing delay.

The energy consumption to serve a request $(u,c)$ consists of the transmission and compute energy. The transmission energy is  
$E^{\text{tx}}(\tau) = \sum_{(u,c)\in\mathcal{R}(\tau)} \sum_{e\in\widetilde{\mathcal{A}}} \varepsilon_e \, b_c \, \mathbb{I}\{e \in \text{path}(u,c,\tau)\}$, where $\varepsilon_e$ is the per-bit energy coefficient of link $e$, and $\text{path}(u,c,\tau)$ is the underlay path used to serve the request $(u,c)$ in slot $\tau$. 
The edge compute energy is
$E^{\text{comp}}(\tau) = \sum_{i\in\mathcal{E}} \sum_{j\in\mathcal{J}} \varepsilon^{\text{comp}}_{i,j}\, x_{i,j,}(\tau)$,
where $\varepsilon^{\text{comp}}_{i,j}$ is the energy cost of producing/converting tile $j$ at edge $i$, and $x_{i,j}(\tau) \in \{0,1\}$ indicates whether node $i$ generates/transcodes tile $j$ in slot $\tau$.

\section{Problem Formulation}
\label{sec:problem}

At epoch $\tau$, each user $u\in\mathcal{U}$ requests
a base segment $m_{u,\tau}\in\mathcal{M}$ together with
the enhancement tiles covering its current FoV.
Let $\mathcal{I}^{\text{Top}}_{u}(\tau)$ and
$\mathcal{I}^{\text{Near}}_{u}(\tau)$ denote the
Top and Near tile-index subsets within the user's viewport,
with $\mathcal{I}^{\mathrm{FoV}}_{u}(\tau)
=\mathcal{I}^{\text{Top}}_{u}(\tau)
\cup\mathcal{I}^{\text{Near}}_{u}(\tau)$.
For each tile index $k\in\mathcal{I}^{\mathrm{FoV}}_{u}(\tau)$
and layer $\ell\in\mathcal{L}$, let
$j(u,\tau,k,\ell)\in\mathcal{J}$ denote the corresponding
tile item.
Each user $u$ is attached to a primary edge
$e(u)\in\mathcal{E}$; the set of neighbor edges
eligible for relay is
$\mathcal{E}^{\mathrm{nbr}}_{e(u)}\subset\mathcal{E}$.
The candidate serving-node set for user $u$ is therefore
$\mathcal{N}_u=\{o\}\cup\{e(u)\}
\cup\mathcal{E}^{\mathrm{nbr}}_{e(u)}$,
where $o$ is the cloud origin.
For the base segment request,
$a_{n,u,m_{u,\tau}}(\tau)\in\{0,1\}$ selects the serving
node $n\in\mathcal{N}_u$.
For each requested FoV tile index
$k\in\mathcal{I}^{\mathrm{FoV}}_{u}(\tau)$, we select both
serving node and layer:
$a_{n,u,k,\ell}(\tau)\in\{0,1\}$ for $n\in\mathcal{N}_u$
and $\ell\in\mathcal{L}$.

A \emph{CDN controller} coordinates edge caching (segments/tiles), compute allocation for tile transcoding, and serving-node selection to maximize a weighted utility balancing hit/relay reward against latency, backhaul usage, energy, and cache placement cost.
It also realizes overlay transfers over the underlay via capacity-aware routing and returns per-request transport costs. Specifically, for delivery of content $c\in\mathcal{C}$ from node $n\in\mathcal{N}_u$ to user $u$ at epoch $\tau$, it provides the completion time ${T}_{n,u,c}(\tau)$ and transmission energy ${E}^{\mathrm{tx}}_{n,u,c}(\tau)$, accounting for path selection, queuing, and fallback attempts.
If a tile is generated at the edge via transcoding, additional processing energy ${E}^{\mathrm{comp}}_{i,j}$ is incurred.

We design a weighted network utility that rewards edge and relay service and penalizes delivery delay, origin traffic, viewport quality mismatch, energy consumption, and cache replacement cost. The weights scalarize the multi-objective design and determine the performance–cost tradeoff.
Let $b_{m_{u,\tau}}$ be the base-segment size,
$b_{k,\ell}$ the size of tile index $k$ delivered at layer $\ell\in\mathcal{L}$,
and $\Delta b_k \triangleq b_{k,\textsf{TOP}}-b_{k,\textsf{NEAR}}\ge 0$.
The indicator function is $\mathbb{I}\{\cdot\}$.
For tile $k$ served from $n\in\mathcal{N}_u$ at layer $\ell\in\mathcal{L}$, per-tile delivery cost is:
\begin{align}
&J_{n,u,k,\ell}(\tau)\triangleq\;
 w_T\, b_{k,\ell}\,{T}_{n,u,j(u,\tau,k,\ell)}(\tau)
+ w_B\, b_{k,\ell}\,\mathbb{I}\{n=o\} \nonumber\\
 & \quad + w_E\, E^{\mathrm{tx}}_{n,u,j(u,\tau,k,\ell)}(\tau) + w_Q\,\Delta b_k\,\mathbb{I}\{k\in\mathcal{I}^{\mathrm{TOP}}_{u}(\tau),\, \nonumber\\
 & \quad\ell=\textsf{NEAR}\} + w_Q\,\Delta b_k\,\mathbb{I}\{k\in\mathcal{I}^{\mathrm{NEAR}}_{u}(\tau),\,\ell=\textsf{TOP}\}.
\label{eq:J_tile_final}
\end{align}
Here $w_T,w_B,w_E,w_Q\ge 0$ are weights; $b_{k,\ell}$ is the delivered tile size (bits);
${T}_{n,u,c}(\tau)$ and ${E}^{\mathrm{tx}}_{n,u,c}(\tau)$ are the completion time and
transmission energy; $\mathbb{I}\{n=o\}$ indicates an origin fetch; and
the last two indicator terms penalize TOP$\rightarrow$NEAR and NEAR$\rightarrow$TOP mismatches,
with $\Delta b_k=b_{k,\textsf{TOP}}-b_{k,\textsf{NEAR}}$.

For the base segment request $m_{u,\tau}$ served from $n\in\mathcal{N}_u$, the delivery cost is:
\begin{equation}
J^{\mathrm{seg}}_{n,u}(\tau)\triangleq
w_B\, b_{m_{u,\tau}}\,\mathbb{I}\{n=o\}
+ w_E\, {E}^{\mathrm{tx}}_{n,u,m_{u,\tau}}(\tau).
\label{eq:J_seg_final}
\end{equation}
where $b_{m_{u,\tau}}$ is the base-segment size, $\mathbb{I}\{n=o\}$ is an origin fetch,
${E}^{\mathrm{tx}}_{n,u,m_{u,\tau}}(\tau)$ is the transmission energy.
The weights $w_B,w_E$ scale the backhaul-traffic and energy penalties.
For tile $k$ served from $n\in\mathcal{N}_u$ at layer $\ell$, we define the Edge-service reward as:
\begin{align}
R^{\mathrm{strict}}_{n,u,k,\ell}(\tau)&\triangleq\;
\mathbb{I}\{n\in\mathcal{E}\}\Big(
b_{k,\textsf{TOP}}\mathbb{I}\{k\in\mathcal{I}^{\mathrm{TOP}}_{u}(\tau),\,\ell=\textsf{TOP}\} \nonumber\\
&+ b_{k,\textsf{NEAR}}\mathbb{I}\{k\in\mathcal{I}^{\mathrm{NEAR}}_{u}(\tau),\,\ell=\textsf{NEAR}\}
\Big). \label{eq:R_strict_final}\\
R^{\mathrm{relay}}_{n,u}(\tau)\triangleq\;
&\mathbb{I}\{n\in\mathcal{E}^{\mathrm{nbr}}_{e(u)}\}. \label{eq:R_relay_final}
\end{align}
where $\mathbb{I}\{n\in\mathcal{E}\}$ indicates edge service (excluding the origin).
The two inner indicators count a strict hit only when a TOP-requested tile is delivered as TOP or a NEAR-requested
tile is delivered as NEAR. The corresponding rewards are bit-weighted via $b_{k,\textsf{TOP}}$ and $b_{k,\textsf{NEAR}}$.
The relay reward $R^{\mathrm{relay}}_{n,u}(\tau)$ is one if the serving node $n$ is a neighbor edge of the primary edge
$e(u)$, i.e., $n\in\mathcal{E}^{\mathrm{nbr}}_{e(u)}$.
To penalize frequent cache replacement, we define eviction indicators
$e^{\mathrm{seg}}_{i,m}(\tau), e^{\mathrm{tile}}_{i,j}(\tau)\in\{0,1\}$ with
$e^{\mathrm{seg}}_{i,m}(\tau)\ge y_{i,m}(\tau-1)-y_{i,m}(\tau),  
e^{\mathrm{tile}}_{i,j}(\tau)\ge y_{i,j}(\tau-1)-y_{i,j}(\tau).$ Finally, the per-epoch utility is:
\begin{align}
&U(\tau)
=\;-\sum_{u,n}
J^{\mathrm{seg}}_{n,u}(\tau)\,
a_{n,u,m_{u,\tau}}(\tau) -w_E \sum_{i,j}
{E}^{\mathrm{comp}}_{i,j}\,x_{i,j}(\tau)
\nonumber\\
&-w_C \sum_{i}
\left(
\sum_{m} e^{\mathrm{seg}}_{i,m}(\tau)
+\sum_{j} e^{\mathrm{tile}}_{i,j}(\tau)
\right)
+\sum_{u,k,\ell,n} a_{n,u,k,\ell}(\tau)\nonumber\\
&\Big(
\eta_{\mathrm{hit}}\,R^{\mathrm{strict}}_{n,u,k,\ell}(\tau)
+\eta_{\mathrm{relay}}\,R^{\mathrm{relay}}_{n,u}(\tau)
 -J_{n,u,k,\ell}(\tau)
\Big).
\label{eq:utility_final_compact}
\end{align}
where $\eta_{\mathrm{hit}},\eta_{\mathrm{relay}},w_C\ge 0$ are also tunable weights.
The first term is the delivery costs for the mandatory base segment (via $J^{\mathrm{seg}}$).
The second and third terms penalize edge-compute energy (via $x_{i,j}(\tau)$) and cache churn (via eviction indicators
$e^{\mathrm{seg}}_{i,m}(\tau)$ and $e^{\mathrm{tile}}_{i,j}(\tau)$), respectively.
The last term rewards strict edge hits and relay service (via $R^{\mathrm{strict}}$ and $R^{\mathrm{relay}}$) and subtracts delivery costs for the FoV tiles (via $J$).
For brevity, the summation $\sum_{u,k,\ell,n}$ is over
$u\in\mathcal{U}$, $k\in\mathcal{I}^{\mathrm{FoV}}_{u}(\tau)$, $\ell\in\mathcal{L}$, and $n\in\mathcal{N}_u$.



We solve a horizon-$T$ discounted utility maximization:
\begin{align}
& \max_{\substack{\mathbf{p}(\tau),\mathbf{y}(\tau),\mathbf{x}(\tau),\\
\mathbf{a}(\tau),\mathbf{e}(\tau)}}
\quad 
\sum_{\tau=1}^{T} \gamma^{\tau-1}\, U(\tau)
\label{eq:opt_main_compact}
\\*                                          
\textrm{s.t.}\quad
&\sum_{m\in\mathcal{M}} y_{i,m}(\tau)\, b_m \le B^{\text{seg}}_{i},
\ \forall i,\tau, \tag{a}\\*                
&\sum_{j\in\mathcal{J}} y_{i,j}(\tau)\, b_j \le B^{\text{tile}}_{i},
\ \forall i,\tau, \tag{b}\\                 
&\sum_{j\in\mathcal{J}} w_j\, x_{i,j}(\tau) \le C^{\mathrm{comp}}_i(\tau),
\ \forall i,\tau, \tag{c}\\
&\sum_{n\in\mathcal{N}_u} a_{n,u,m_{u,\tau}}(\tau) \le 1,
\ \forall u,\tau, \tag{d}\\*               
&\sum_{n\in\mathcal{N}_u}\sum_{\ell\in\mathcal{L}} a_{n,u,k,\ell}(\tau) \le 1,
\ \forall u,k,\tau, \tag{e}\\              
&a_{i,u,m_{u,\tau}}(\tau)\le y_{i,m_{u,\tau}}(\tau),
\ \forall i,u,\tau, \tag{f}\\*             
&a_{i,u,k,\ell}(\tau)
\le y_{i,j(u,\tau,k,\ell)}(\tau)
+x_{i,j(u,\tau,k,\ell)}(\tau), \nonumber\\*
&\hspace{2.2cm}\forall i,u,k,\ell,\tau, \tag{g}\\  
&e^{\mathrm{seg}}_{i,m}(\tau)\ge y_{i,m}(\tau\!-\!1)-y_{i,m}(\tau),
\ \forall i,m,\tau, \tag{h}\\*             
&e^{\mathrm{tile}}_{i,j}(\tau)\ge y_{i,j}(\tau\!-\!1)-y_{i,j}(\tau),
\ \forall i,j,\tau. \tag{i}
\end{align}
where $\mathbf{p}(\tau)=\{p_{i,c}(\tau)\}_{i\in\mathcal{E},\,c\in\mathcal{C}}$,
$\mathbf{y}(\tau)=\{y_{i,m}(\tau),y_{i,j}(\tau)\}_{i\in\mathcal{E},\,m\in\mathcal{M},\,j\in\mathcal{J}}$,
$\mathbf{x}(\tau)=\{x_{i,j}(\tau)\}_{i\in\mathcal{E},\,j\in\mathcal{J}}$,
$\mathbf{a}(\tau)=\{a_{n,u,m_{u,\tau}}(\tau),a_{n,u,k,\ell}(\tau)\}_{u\in\mathcal{U},\,n\in\mathcal{N}_u,\,k\in\mathcal{I}^{\mathrm{FoV}}_{u}(\tau),\,\ell\in\mathcal{L}}$,
and $\mathbf{e}(\tau)=\{e^{\mathrm{seg}}_{i,m}(\tau),e^{\mathrm{tile}}_{i,j}(\tau)\}_{i\in\mathcal{E},\,m\in\mathcal{M},\,j\in\mathcal{J}}$.
Constraints~\eqref{eq:opt_main_compact}(a)--(b) enforce the per-edge storage capacity for the segment cache and tile cache, respectively. Constraint~\eqref{eq:opt_main_compact}(c) limits edge compute dispatch within the per-epoch available compute resources. Constraint~\eqref{eq:opt_main_compact}(d) selects exactly one serving node for each base-segment request, while~\eqref{eq:opt_main_compact}(e) selects exactly one serving pair for each requested FoV tile. Constraints~\eqref{eq:opt_main_compact}(f)--(g) ensure service feasibility at the edge: a segment can be served by an edge only if it is cached, and a tile can be served only if it is cached or produced via edge compute. Finally,~\eqref{eq:opt_main_compact}(h)--(i) define eviction indicators used to account for cache churn.

\section{Problem Reformulation}
\label{sec:reform}

The optimization problem in~\eqref{eq:opt_main_compact} is difficult to solve online due to combinatorial cache decisions under storage constraints (a)--(b), binary compute dispatch $x_{i,j}(\tau)$ coupled through edge compute allocation~(c), and time-varying network conditions induced by shared-link congestion and relay traffic.
To enable real-time deployment, we reformulate it as a decentralized sequential decision problem at the edge, where cloud pre-distribution is specified by a fixed planner $p_{i,c}(\tau)=\pi^{\mathrm{cloud}}(i,c;\tau)$ and request delivery is handled by a priority-based serving policy coupled with a capacity-aware network performance estimator $\mathcal{M}^{\mathrm{net}}(\tau)$.
Rather than optimizing the service variables $\mathbf{a}(\tau)$ directly, we let the serving policy determine the effective service outcome as $\mathbf{a}(\tau)=\mathcal{F}_{\mathrm{serve}}\!\left(\mathbf{y}(\tau),\mathbf{x}(\tau),\pi^{\mathrm{cloud}}(\tau),\mathcal{M}^{\mathrm{net}}(\tau)\right)$, where $\mathcal{M}^{\mathrm{net}}(\tau)$ returns $\big({T}_{n,u,c}(\tau),\,{E}^{\mathrm{tx}}_{n,u,c}(\tau)\big)$ capturing capacity-aware path selection, shared-link queuing, and failover.
The online control problem therefore reduces to sequential edge decisions over $\big(\mathbf{y}(\tau),\mathbf{x}(\tau)\big)$ under partial observability.

\subsection{Markov Decision Process Formulation}
We model the system as a Markov Decision Process (MDP) 
$\langle \mathcal{S},\{\mathcal{O}_i\}_{i\in\mathcal{E}},\{\mathcal{A}_i\}_{i\in\mathcal{E}},\mathcal{P},\{\mathcal{R}_i\}_{i\in\mathcal{E}},\gamma\rangle$.
%
At each epoch $\tau$, each edge $i\in\mathcal{E}$ receives a local observation $O_{i,\tau}\in\mathcal{O}_i$ and selects an action
$A_{i,\tau}\in\mathcal{A}_i$. The environment then evolves according to the transition probability $\mathcal{P}$, which captures cloud pre-distribution under $\pi^{\mathrm{cloud}}$, edge cache/compute updates, and request service realized by the priority-based serving policy together with the capacity-aware network performance estimator.
The resulting per-edge rewards $\{\mathcal{R}_i(\tau)\}$ are returned to the agents, and the process repeats.

\paragraph{State}
The global state $S_\tau$ includes:
per-edge segment and tile cache states (including any eviction-protection metadata),
per-edge compute queues and remaining compute resources, 
 user--edge associations and the realized request, and
 remaining capacities and in-flight transfers that determine the queuing delay
$T^{\mathrm{queue}}$ in~\eqref{eq:transfer_completion_time}.
Because each edge observes only local cache/compute states and limited network feedback, the resulting decision problem is partially observable.
Each edge $i$ forms a fixed-dimensional observation vector from locally available states with observation dimension $d_o$:
\begin{equation}
O_{i,\tau}=
\big[
\mathbf{o}^{\mathrm{stat}}_{i,\tau};\
\mathbf{o}^{\mathrm{tail}}_{i,\tau};\
\mathbf{o}^{\mathrm{topk}}_{i,\tau}
\big]\in\mathbb{R}^{d_o},
\label{eq:obs_316}
\end{equation}
The three blocks summarize: (i) local states (cache occupancy, in-flight transfers, compute load, and recent hit/churn signals),
(ii) tail-demand statistics capturing mass beyond the locally tracked popular set, and
(iii) bounded-dimensional Top-$K$ demand features that encode content popularity, size, and transport-cost proxies across multiple granularities.

\paragraph{Continuous action parameterization}
Direct optimization over binary keep/evict decisions is combinatorial and thus impractical for online decisions.
We instead parameterize each edge action as a five-dimensional continuous vector
\begin{equation}
\mathbf{a}_{i}(\tau)
=\big(\beta_i,\,u_i^{\mathrm{pop}},\,u_i^{\mathrm{size}},\,u_i^{\mathrm{cost}},\,u_i^{\tau}\big)
\in(0,1)^5,
\label{eq:act_5d}
\end{equation}
sampled from a stochastic policy $\pi_i(\cdot\mid O_{i,\tau})$ using independent Beta distributions.
The raw outputs are mapped to actionable controls by
$w_i^{\mathrm{pop}} = W_{\max}\,(2u_i^{\mathrm{pop}}-1)$,
$w_i^{\mathrm{size}} = W_{\max}\,(2u_i^{\mathrm{size}}-1)$,
$w_i^{\mathrm{cost}} = W_{\max}\,(2u_i^{\mathrm{cost}}-1)$,
$\tau_i = T_{\max}\,(2u_i^{\tau}-1)$, and
$C_i^{\mathrm{comp}}(\tau) = \beta_i\,\overline{C}_i^{\mathrm{comp}}$,
where $W_{\max}$ and $T_{\max}$ are design constants.
Here $\big(w_{i,\tau}^{\mathrm{pop}},w_{i,\tau}^{\mathrm{size}},w_{i,\tau}^{\mathrm{cost}}\big)$ are cache-admission weights that
respectively promote popular objects, penalize large objects, and favor objects with higher expected transport-cost savings.
The threshold $\theta_{i,\tau}$ controls overall admission selectivity, while $\beta_{i,\tau}$ allocates a fraction of the peak compute
capacity $\overline{C}_i^{\mathrm{comp}}$ to tile processing.


\paragraph{Action-to-cache/compute mapping}
At epoch $\tau$, edge $i$ forms segment/tile candidate pools for cache update:
$\mathcal{T}^{\text{seg}}_{i}(\tau) := \mathcal{C}^{\text{seg}}_{i}(\tau\!-\!1)\cup \mathcal{A}^{\text{seg}}_{i}(\tau)$,
$\mathcal{T}^{\text{tile}}_{i}(\tau) := \mathcal{C}^{\text{tile}}_{i}(\tau\!-\!1)\cup \mathcal{A}^{\text{tile}}_{i}(\tau)$,
where $\mathcal{A}_{i}(\tau)$ are newly arrived contents (cloud push, relay pull, cache fill, etc.).
Each candidate content $c$ is assigned a local admission score: 
\begin{equation}
\mathrm{score}_i(c;\tau)=
w_i^{\mathrm{pop}}\,\widetilde{\mathrm{pop}}_i(c)
- w_i^{\mathrm{size}}\,\widetilde{\mathrm{size}}(c)
+ w_i^{\mathrm{cost}}\,\widetilde{\mathrm{cost}}_i(c),
\label{eq:admit_score_impl}
\end{equation}
where $\widetilde{\mathrm{pop}}_i$, $\widetilde{\mathrm{size}}$, and
$\widetilde{\mathrm{cost}}_i$ are locally computed normalized signals capturing (estimated) popularity, object size, and an expected transport-cost saving by caching.
Edge $i$ admits items greedily in descending score order within each cache
if $\mathrm{score}_i(c;\tau)\ge\max\{0,\,\tau_i\}$
and capacity constraints (\ref{eq:opt_main_compact}.a)/(\ref{eq:opt_main_compact}.b) remain satisfied,
producing $\mathcal{C}^{\text{seg}}_{i}(\tau)$ and $\mathcal{C}^{\text{tile}}_{i}(\tau)$,
equivalently $y_{i,m}(\tau)$ and $y_{i,j}(\tau)$. 
Compute scheduling decisions satisfy the per-epoch compute resources~(\ref{eq:opt_main_compact}.c) and are further restricted by service feasibility constraints (e.g., compute-pressure, latency, and quality requirements), as described in Section~\ref{sec:problem}.

\subsection{Reward Reformulation}
\label{subsec:reward_reform}

To enable decentralized learning, we decompose the global objective~\eqref{eq:utility_final_compact} into rewards per edge server by attributing request-related terms to the user’s allocated edge, while keeping energy and cache replacement costs per edge.
This decomposition preserves the original objective, $
\sum_{i\in\mathcal{E}} R_i(\tau) = U(\tau)$
under the chosen attribution rule, while providing each edge with a local reward.
Therefore, maximizing the discounted return:
\begin{equation}
\max_{\{\pi_i\}_{i\in\mathcal{E}}}\;
\mathbb{E}\!\left[\sum_{\tau=1}^{T}\gamma^{\tau-1}\sum_{i\in\mathcal{E}} R_i(\tau)\right]
\label{eq:rl_obj}
\end{equation}
optimizes the same QoE--cost trade-off targeted by~\eqref{eq:opt_main_compact}, while enabling online, decentralized execution.

The resulting reformulation yields a learning framework where at each epoch $\tau$, edge $i$ samples its control action from
$\pi_i(\mathbf{a}_i(\tau)\!\mid\!O_{i,\tau})$ using only local observations. The environment---including cloud pre-distribution, edge cache/compute updates,
the priority-based serving policy, and the capacity-aware network performance estimator---then realizes the resulting deliveries and returns
the corresponding per-edge rewards. The next section instantiates $\{\pi_i\}$ using PPO under centralized training and decentralized execution.

\subsection{Training and Execution Workflow}
\label{subsec:train_exec}

We model training as episodic interaction with the environment. At the start of each episode, the CDN environment is reset, including edge caches, compute queues, and in-flight transfers, and then evolves over $H$ epochs, each aligned with one segment window. At epoch $\tau$, the cloud planner applies the FoV-aware pre-distribution policy $\pi^{\mathrm{cloud}}$, aggregates per-edge viewport statistics, identifies the Top and Near tile sets, and pushes the selected base segments and enhancement tiles to the corresponding edges with a lookahead of $L$ segments. Each edge $i$ observes its local state $O_{i,\tau}$ in~\eqref{eq:obs_316} and samples a continuous action $\mathbf{a}_i(\tau)\sim\pi_{\theta_i}(\cdot\mid O_{i,\tau})$ as parameterized in~\eqref{eq:act_5d}. The action is mapped to cache-admission weights and a compute-budget fraction, which determine cache updates via the greedy rule in~\eqref{eq:admit_score_impl} under constraints~(\ref{eq:opt_main_compact}a)--(\ref{eq:opt_main_compact}b), and compute dispatch under~(\ref{eq:opt_main_compact}c).

Requests are then served in the fixed priority order \emph{cache hit} $\rightarrow$ \emph{edge compute hit} $\rightarrow$ \emph{relay hit} $\rightarrow$ \emph{origin fetch}. For relay candidates, the network performance estimator returns ${T}_{n,u,c}(\tau)$ and ${E}^{\mathrm{tx}}_{n,u,c}(\tau)$, and the relay source with the lowest delivery cost is selected. The resulting per-edge rewards $\{R_i(\tau)\}$ are computed from the decomposed utility and stored with the corresponding transitions. After each episode, advantages are estimated using Generalized Advantage Estimation (GAE). Each actor $\pi_{\theta_i}$ is updated by maximizing the PPO clipped surrogate objective:
$\mathcal{L}_{\pi_i}
=
\mathbb{E}\!\left[
\min\!\left(r_t\hat{A}_{i,t},\ \mathrm{clip}(r_t,1\!-\!\epsilon,1\!+\!\epsilon)\hat{A}_{i,t}\right)
\right]
+\alpha\,\mathcal{H}(\pi_i),
\label{eq:ppo_actor_obj}$
where $r_t = \pi_{\theta_i}(a_{i,t} \mid o_{i,t}) / \pi_{\theta_i}^{\mathrm{old}}(a_{i,t} \mid o_{i,t})$ is the importance ratio and $\alpha\,\mathcal{H}(\pi_i)$ is the entropy bonus. The centralized critic is trained by minimizing
$\mathcal{L}_{V}
=
\mathbb{E}\!\left[\left(V_\phi(s_t) - \hat{R}_t\right)^2\right].
\label{eq:ppo_critic_obj}$
At deployment, the critic is discarded and each edge executes its learned policy $\pi_{\theta_i}(\cdot \mid O_{i,\tau})$ using only local observations. The full procedure is summarized in Algorithm~\ref{alg:ctde_mappo_impl}.

\begin{algorithm}[t]
\caption{ImCDN-MAPPO: Immersive CDN with MAPPO}
\label{alg:ctde_mappo_impl}
\begin{algorithmic}[1]
\REQUIRE Horizon $H$, lookahead $L$, Top budget $M$, cache caps $(B^{\mathrm{seg}}_i,B^{\mathrm{tile}}_i)$,
compute cap $\overline{C}^{\mathrm{comp}}_i$, cloud planner $\pi^{\mathrm{cloud}}$, estimator $\mathcal{M}^{\mathrm{net}}$.
\ENSURE Trained decentralized actors $\{\pi_{\theta_i}\}_{i\in\mathcal{E}}$.

\STATE Initialize actors $\{\pi_{\theta_i}\}$, critic $V_\phi$, buffer $\mathcal{B}\!\leftarrow\!\emptyset$.
\FOR{episode $=1,\dots,E_{\max}$}
    \STATE Reset environment, caches, in-flight state.
    \FOR{$\tau=1,\dots,H$}
        \STATE $\{\mathcal{A}^{\mathrm{seg}}_i,\mathcal{A}^{\mathrm{tile}}_i\}_{i\in\mathcal{E}}\!\leftarrow\!\pi^{\mathrm{cloud}}(\tau;L,M)$.
        \FORALL{edge $i\in\mathcal{E}$ \textbf{in parallel}}
            \STATE Observe $O_{i,\tau}$; sample $\mathbf{a}_i\!\sim\!\pi_{\theta_i}(\cdot|O_{i,\tau})$.
            \STATE Score via~(\ref{eq:admit_score_impl}) \& admit cache; dispatch compute under~(\ref{eq:opt_main_compact}a)--(\ref{eq:opt_main_compact}c).
        \ENDFOR
        \STATE Serve: \emph{hit\,$\to$\,compute\,$\to$\,relay\,$\to$\,origin}; query $\mathcal{M}^{\mathrm{net}}$; commit transfers.
        \STATE  Compute $\{\mathcal{R}_i\}_{i\in\mathcal{E}}$; Store $(O_{i,\tau},\mathbf{a}_i,R_i,O_{i,\tau\!+\!1})$ in $\mathcal{B}$.
    \ENDFOR
    \STATE GAE $\to$ update $\{\pi_{\theta_i}\}$, $V_\phi$ via~\eqref{eq:ppo_actor_obj}; clear $\mathcal{B}$.
\ENDFOR
\STATE \textbf{Deploy:} discard $V_\phi$; each edge runs $\pi_{\theta_i}(\cdot\mid O_{i,\tau})$ locally.
\end{algorithmic}
\end{algorithm}

\section{Simulation Results}
The CDN topology consists of $1$ cloud nodes, $5$ edge nodes, $50$ relay nodes,
and $150$ user nodes over a $100 \times 100$~km$^{2}$ area.
Each 360$^\circ$ video is tiled on a $12 \times 6$ equirectangular
grid with segment duration $\Delta t_{\mathrm{seg}} = 1$~s.
Five videos from the 360VidStr dataset \cite{360vidstr_dataset_github_2021} are used with traced
user head orientations. We use a horizon of $H = 30$ steps per episode, with $\mathrm{cap}_e(\tau) = C_e \cdot 10^6 \cdot \Delta t_{\mathrm{seg}}$ as the per-epoch bit budget on link $e$. The weights in the utility are: $\eta_{\mathrm{hit}}=\eta_{\mathrm{relay}}=16.0$, $w_T=2.5$, $w_B=0.012$, $w_E$ (tx / comp)$=0.012$ / $0.03$, $w_Q=1.2$, $w_C=0.002$. $\kappa=0.005$ (ms/km), $\delta_0^{\text{bb}}=1.5$~ms,
$\delta_0^{\text{ac}}=5.0$~ms, $\delta_{\min}^{\text{ac}}=0.5$~ms, $\sigma_{\text{ac}}=1.5$, $\mathcal{T}_{\mathrm{ac}}=\{100,300,500,1000\}$~Mbps,
$\zeta_{\min}=0.5$, $\zeta_{\max}=1.0$, $X_{\min}=5$, $X_{\max}=10$, $\lambda=10^3$,
$C_{\min}^{\mathrm{ac}}=20$~Mbps, and $C_{\max}^{\mathrm{ac}}=10^3$~Mbps. Other parameters are listed
in Table~\ref{tab:sim_params}.

\begin{table}[t]
\centering
\caption{Simulation Parameters}
\label{tab:sim_params}
\small
\begin{tabularx}{\linewidth}{Xr}
\toprule
\textbf{Parameter} & \textbf{Value} \\
\midrule
Backbone $k$-NN connectivity & $k = 5$ \\
User--relay access links & $5$ nearest \\
Base segment bitrate (2K) & $12$~Mbps \\
Top tile enhancement (6K) & $1{,}200$~kbits/tile \\
Near tile enhancement (4K) & $800$~kbits/tile \\
Viewport FoV $(\theta_h \times \theta_v)$ & $100^\circ \times 100^\circ$ \\
Top-tile budget $M$ & $8$ \\
Per-edge segment cache $B^{\mathrm{seg}}_i$ & $40$~MB \\
Per-edge tile cache $B^{\mathrm{tile}}_i$ & $80$~MB \\
Edge transcode budget $C^{\mathrm{comp}}_i$ & $35$~Mbits/epoch \\
Cache retention window & $4$ segments \\
Discount factor $\gamma$ & $0.99$ \\
GAE parameter $\chi$ & $0.95$ \\
PPO clip ratio $\epsilon$ & $0.2$ \\
PPO update epochs $K$ & $4$ \\
Actor / Critic learning rate & $3 \times 10^{-4}$ \\
Action dimension & $5$ \\
Observation dimension $d_o$ & $316$ \\
\bottomrule
\end{tabularx}
\end{table}


\begin{figure}
    \centering
    \includegraphics[width=1\linewidth]{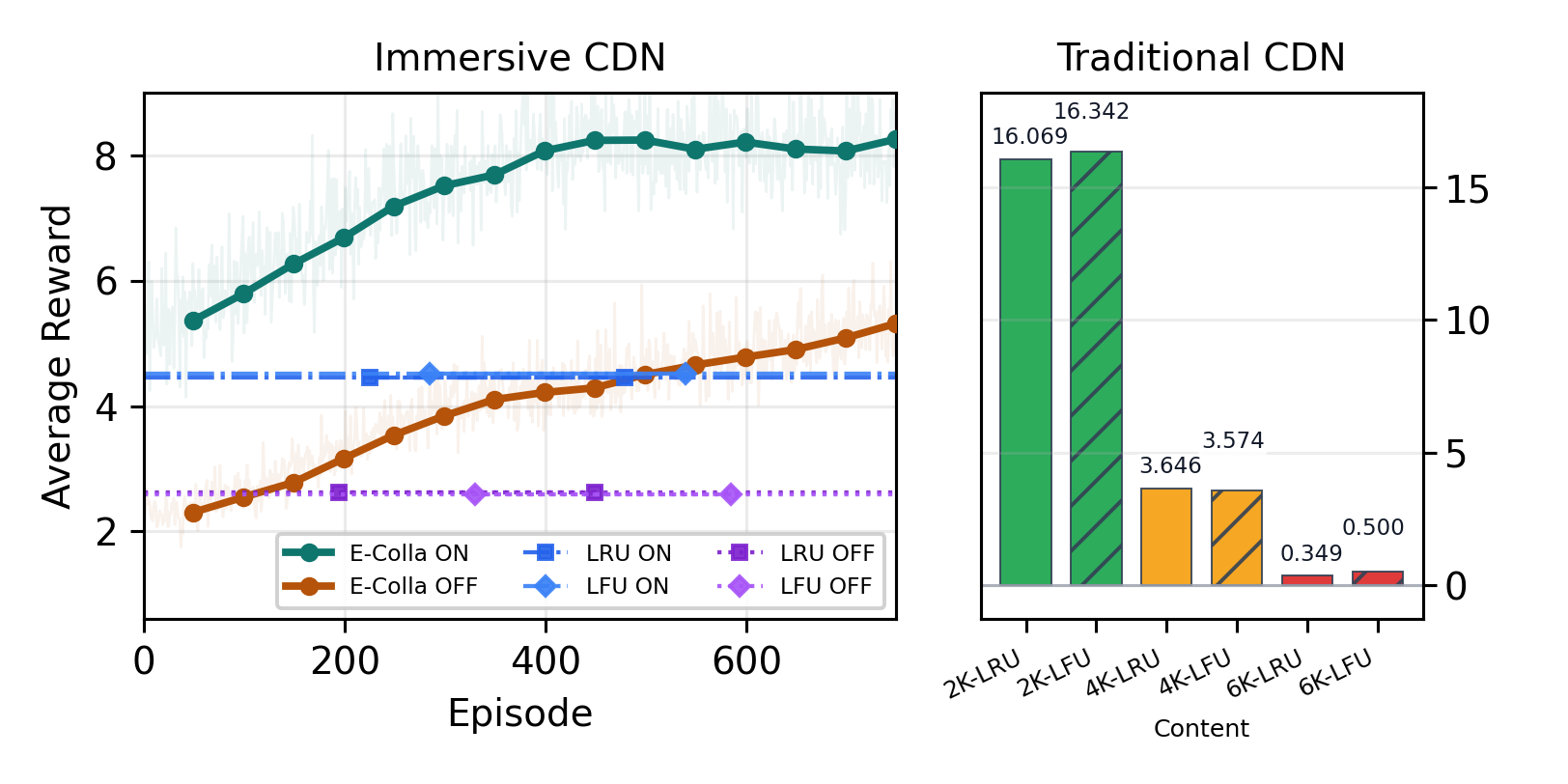}
    \caption{Average Reward ImCDN (left) and traditional CDN (right).}
    \label{fig:Average Reward}
\end{figure}

Fig.~\ref{fig:Average Reward} shows the reward of the proposed ImCDN with the MAPPO-based algorithm. In the left subplot, the MAPPO-based policy with edge cooperation (green) nearly doubles the reward of its non-collaborative counterpart (brown) and consistently outperforms the LRU and LFU baselines in both cases. In the right subplot, we evaluate LRU and LFU under a traditional CDN, where only full segments at a single rendering rate are delivered. \textcolor{black}{
Transmitting only 2K base segments keeps per-segment size small
and thus achieves a higher cache hit rate with a higher reward, but fails to satisfy higher immersive quality requirements.
On the other hand, transmitting full 4K or 6K segments satisfies
quality demands but dramatically increases delivery cost and
backhaul traffic, leading to a substantial drop in overall reward.
}


\begin{figure}
    \centering
    \includegraphics[width=1\linewidth]{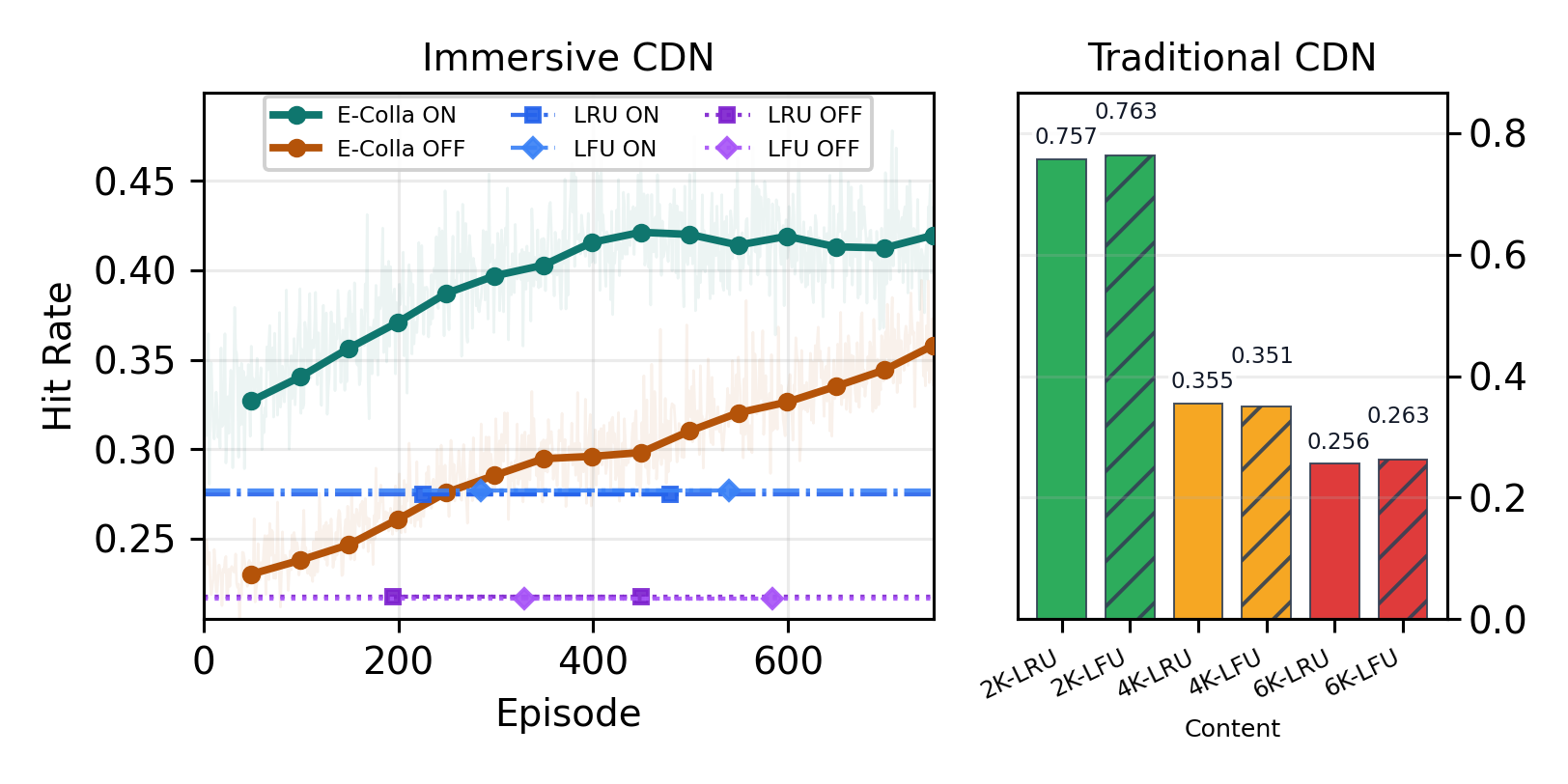}
    \caption{Hit Rate for ImCDN (left) and traditional CDN (right).}
    \label{fig:Hit Rate}
\end{figure}

Fig.~\ref{fig:Hit Rate} compares the pre-caching hit rates achieved by each method. The overall trend is consistent with the reward. The ImCDN powered by the MAPPO-based algorithm with edge cooperation achieves a hit rate exceeding 40\%, outperforming both the non-collaborative RL variant and the LRU/LFU baselines. In traditional CDNs, caching only  full segments with high rendering rate lowers the overall hit rate because they consume more cache space under limited storage. In contrast, our ImCDN jointly caches both segments and tiles, enabling fine-grained tile-level storage that saves cache space, which achieves a hit rate superior to that of caching 4K full segments.

\begin{figure}
    \centering
    \includegraphics[width=1\linewidth]{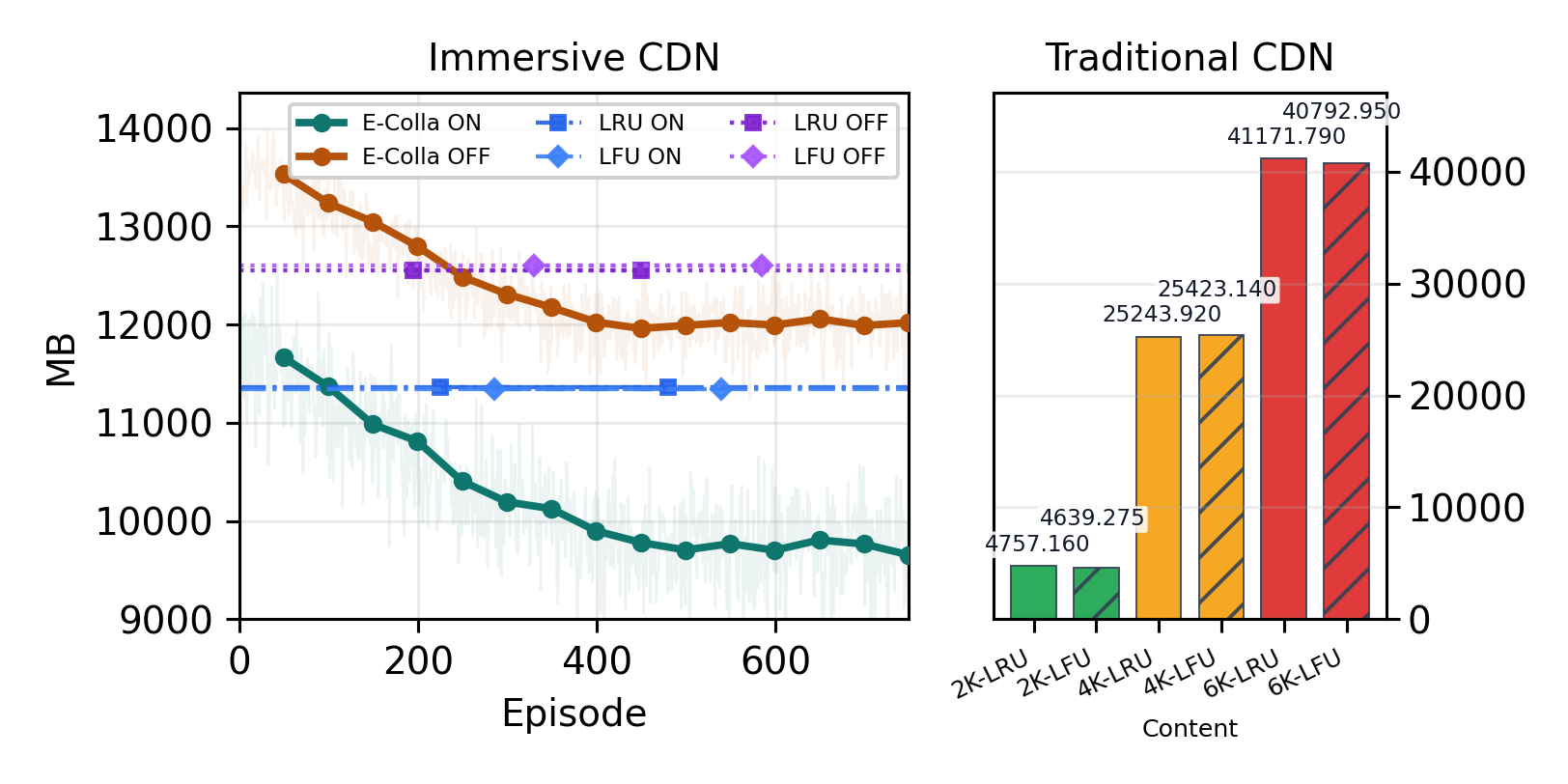}
    \caption{Origin Traffic for ImCDN (left) and traditional CDN (right).}
    \label{fig:Origin Traffic}
\end{figure}

Fig.~\ref{fig:Origin Traffic} illustrates the origin traffic volume when cache misses occur and content must be fetched from the cloud. Thanks to its higher hit rate, the ImCDN with MAPPO and edge collaboration achieves the lowest origin traffic, reducing it by approximately 18\% compared to the non-collaborative RL variant and outperforming the conventional LRU/LFU baselines as well. Traditional CDNs reduce origin traffic by delivering only 2K base segments, but this cannot meet users' higher rendering-rate demands. By contrast, our ImCDN enables higher rendering quality while greatly reducing origin traffic relative to 4K or 6K full-segment delivery.

\begin{table}[t]
\centering
\caption{Metric comparison (Best Performance)}
\label{tab:metric_comparison_clean_tabx}
\small
\begin{tabularx}{\linewidth}{Xrrrr}
\toprule
Configuration & Reward & HitRate & Wait (s) & Origin (MB) \\
\midrule
Im-MA (On) & 9.850 & 0.478 & 0.197 & 8479.6 \\
Im-MA (Off) & 6.310 & 0.394 & 0.241 & 11363.0 \\
Im-LRU (On) & 4.462 & 0.275 & 0.221 & 11359.3 \\
Im-LFU (On) & 4.511 & 0.277 & 0.220 & 11344.3 \\
Im-LRU (Off) & 2.620 & 0.218 & 0.249 & 12552.8 \\
Im-LFU (Off) & 2.596 & 0.217 & 0.249 & 12603.7 \\
Tr-LRU (2K) & 16.069 & 0.757 & 0.277 & 4757.2 \\
Tr-LFU (2K) & 16.342 & 0.763 & 0.276 & 4639.3 \\
Tr-LRU (4K) & 3.646 & 0.355 & 0.546 & 25243.9 \\
Tr-LFU (4K) & 3.574 & 0.351 & 0.551 & 25423.1 \\
Tr-LRU (6K) & 0.349 & 0.256 & 0.596 & 41171.8 \\
Tr-LFU (6K) & 0.500 & 0.263 & 0.594 & 40792.9 \\
\bottomrule
\end{tabularx}
\end{table}

 
Table~\ref{tab:metric_comparison_clean_tabx} summarizes the best observed performance of each algorithm across several key metrics. Consistent with the preceding analysis, the proposed ImCDN with MAPPO (Im-MA) and edge collaboration achieves the highest reward, the highest hit rate, and the lowest origin traffic among all configurations---excluding the traditional case of transmitting only base segments (Tr-LRU/LFU). More notably, it also attains the lowest average user waiting time compared to all other delivery configurations. This demonstrates that our approach can better accommodate users' diverse viewport and rendering-rate requirements while still maintaining the best overall latency performance. 

\section{Conclusion and Future Work}
\label{sec:conclusion}

This paper presented ImCDN, an end-to-end, tile-level, FoV-aware framework for 360° video delivery that integrates FoV-driven cloud pre-distribution, hybrid segment–tile caching at edge servers, limited edge transcoding capacity, and edge collaboration. We formulated joint CDN optimization as a utility-maximization problem and cast it as an MDP solved by MAPPO. 
Experiments on a realistic topology with real head-tracking traces show that ImCDN outperforms segment-only CDNs at comparable immersive quality levels (e.g., 4K/6K) while achieving the lowest waiting time among all tested configurations. It also outperforms standard caching heuristics, while edge collaboration nearly doubles the reward. 
In future work, we will jointly optimize cloud pre-distribution with edge cache admission.

\bibliographystyle{ieeetr}
\bibliography{Ning.bib}

\end{document}